\documentclass[aps,pra,preprint,groupedaddress]{revtex4-2}

\usepackage{multirow}%
\usepackage{amsmath,amssymb,amsfonts}%
\usepackage{amsthm}%
\usepackage{bm}
\usepackage{amsmath}
\usepackage{mathrsfs}%
\usepackage{graphicx}%

\begin{document}

\preprint{}

\title{Measurement of the total spin angular momentum $\langle F_z \rangle$ of alkali-metal atoms}


\author{Runa Yasuda}\affiliation{Department of Applied Physics, Tokyo University of Agriculture and Technology, Koganei, Tokyo 184-8588, Japan}
\author{Kei Ishii}\affiliation{Department of Applied Physics, Tokyo University of Agriculture and Technology, Koganei, Tokyo 184-8588, Japan}
\author{Wolfgang Klassen}  \affiliation{Department of Physics and Astronomy, University of Manitoba, Winnipeg, MB R3T 2N2 Canada}
\affiliation{Department of Physics and Astronomy, The University of British Columbia, Vancouver, BC V6T 1Z1, Canada}\author{Jeffery W. Martin} \affiliation{Department of Physics, The University of Winnipeg, Winnipeg, MB R3B 2E9, Canada}
\author{Atsushi Hatakeyama} \email{hatakeya@cc.tuat.ac.jp}
\affiliation{Department of Applied Physics, Tokyo University of Agriculture and Technology, Koganei, Tokyo 184-8588, Japan}


\date{\today}

\begin{abstract}
It is important to evaluate the total spin angular momentum of alkali-metal atoms if the atoms serve as a reservoir of angular momenta.
We use an absorption-monitoring technique to measure $\langle F_z \rangle$, i.e., the expectation values of the quantization ($z$) axis components of the total angular momentum of cesium (Cs) atoms in the electronic ground state in both uncoated and anti-relaxation-coated vacuum cells at room temperature. Cs atoms are polarized via optical pumping and probed using their $D_{2}$ transitions. The probe laser frequency is varied across the Doppler-broadened $D_2$ transition; the $\langle F_z \rangle$ values are derived using the integrated absorption coefficients. 
The largest $\langle F_z \rangle$ is 2.5 for the coated cell. We then use a simple model of spin flow through vapor cells to estimate the atomic spin relaxation probabilities after a single surface collision.
\end{abstract}


\maketitle

\section{Introduction}\label{sec:intro}
Spin-polarized gaseous atoms, especially alkali-metal vapors, play key roles in many experiments conducted in the field of atomic physics. The electronic ground states of alkali-metal atoms exhibit three types of spin: electronic spin $\bm{S}$, nuclear spin $\bm{I}$, and total spin $\bm{F}=\bm{S}+\bm{I}$. The electronic spin magnitude is $S=1/2$, attributable to the outermost single electron, which lacks orbital angular momentum. The associated magnetic moment is larger (by three orders of magnitude) than the nuclear magnetic moment. Thus, electronic spin is sensitive to external perturbations, including magnetic fields, and to collisions with other atoms and container walls. Electronic spin thus plays important roles during the spin-exchange optical pumping of noble gas nuclei \cite{Wal97, App98}, and affects the magnetometric properties of spin-exchange-relaxation-free (SERF) regimes \cite{All02, Kom03}. 

The nuclear spins of stable alkali-metal atoms are larger than the electronic spin. The smallest is $I = 1$ for $^6$Li, and the largest is $I = 7/2$ for $^{133}$Cs. The nuclear and electronic spins are coupled via hyperfine interactions. In the term that expresses the resulting total spin $\bm{F}$, $\bm{F}^2$ and $F_z$ are constants of motion in a weak magnetic field, $B \ll A_{\rm{HF}} / \mu_{B}$, where $B$ is the magnetic flux density along the $z$ axis, $A_{\rm{HF}}$ is the hyperfine structure constant ($h\times$2.3~GHz for the 6$^2S_{1/2}$ level and $h\times$50~MHz for the 6$^2P_{3/2}$ level of Cs atoms\cite{Ari77, Ger03} ($h$ is Planck's constant)), and $\mu_{B}$ is the Bohr magneton. $A_{\rm{HF}} / \mu_{B} \sim 4$~mT, and $\sim$ 0.2~T for 6$^2P_{3/2}$ and 6$^2S_{1/2}$ levels of the Cs atom, respectively. Given the large magnitudes thereof, the nuclear spins store most of the angular momentum of ground-state alkali-metal atoms; the nuclei thus serve as a ``flywheel'' \cite{Bal75}.

The extent of spin polarization is often described by using the expectation values of the spin component along the $z$ axis, i.e., $\langle S_z\rangle$, $\langle I_z\rangle$, and $\langle F_z\rangle$. These facilitate an understanding of the spin dynamics of optically polarized atoms \cite{Bal75}. 
High-level spin polarization is often experimentally associated with a large signal-to-noise ratio; this is of concern, particularly when the number of spins is very small \cite{Nis19}. 

Since the 1950s, alkali-metal atom polarizations have been measured using various methods. Although the nuclear polarizations of certain unstable nuclei can be derived by using $\beta$-decay asymmetry at the level of a single nucleus \cite{Lev03}, most methods employ the spin-dependent optical properties of polarized samples. The optical responses of spin-polarized atoms depend, in a complicated manner, on the polarization moments of the ground-state alkali-metal atoms \cite{Hap72, Auz10}, but are affected by only $\langle S_z\rangle$ under the following conditions: the hyperfine levels of alkali-metal atom ground states cannot be optically resolved if either the pressure broadenings caused by the buffer gas are larger than the hyperfine intervals \cite{Wal97} or the light linewidth is excessively broad \cite{Bou65}. This is often the case when seeking to study the spin-exchange optical pumping of noble gas nuclei. Under such conditions, absorption or fluorescence monitoring using a circularly polarized probe light can be employed to measure $\langle S_z\rangle$ values, which are proportional to the probe light absorption.

Absorption or fluorescence monitoring is sometimes inapplicable; for example, when studying optically dense vapors that transmit light poorly, fluorescent light is readily saturated. In such cases, birefringence monitoring is useful \cite{Str67, Mat70, Led08, LiY17}. This is based on the difference between the refractive indices of right- and left-circularly polarized light, which rotates the polarization plane of the incident linearly polarized light. Off-resonance light absorption is negligible but the difference in the refractive index remains observable; this approach is very useful in optically thick conditions such as SERF scenarios. The birefringence signal of off-resonant light is also proportional to $\langle S_z \rangle$ \cite{Str67, Ros07, Din16}. 

Although many $\langle S_z\rangle$ values are obtained, $\langle I_z \rangle$ and $\langle F_z \rangle$ data are scarce. The $\langle I_z \rangle$ parameter is important when studying nuclear physics \cite{Nis19}, including nuclear magnetic resonance (NMR). The $\langle F_z \rangle$ parameter enhances the understanding of the spin dynamics during optical pumping, including spin relaxation \cite{Bal75}. If polarized gas atoms are viewed as a reservoir of angular momenta \cite{Hat19}, the evaluation of $\langle F_z \rangle$ is particularly important.
In one work, $\langle F_z \rangle$ was estimated by using birefringence monitoring at an assumed spin temperature \cite{Str67}. In principle, $\langle S_z \rangle$, $\langle I_z \rangle$, and $\langle F_z \rangle$ can be calculated when the populations of all magnetic sublevels are known. This is particularly useful during laser-cooling or high-magnetic-field experiments; the Zeeman sublevels are then optically resolved \cite{Ols11,Ste11}. When studying hot alkali-metal vapors in low fields, the populations of the magnetic sublevels were reconstructed using the differences between the optical transition strengths of the Zeeman sublevels \cite{Lon10}. However, the populations of the ground-state sublevels were not completely evaluated, and $\langle F_z \rangle$ could thus not be derived.

Here, we use absorption monitoring to measure the $\langle F_z \rangle$ parameters of the total cesium (Cs) spin angular momenta of uncoated and anti-relaxation-coated vapor cells at room temperature under a low magnetic field. The method exploits the differences between the optical transition strengths of the various magnetic sublevels and requires only well-resolved hyperfine levels of the electronic ground state. The applications are thus broader than described in Ref. \cite{Str67}. We polarized Cs vapors via optical pumping by using the $D_{2}$ transition and probed $\langle F_{z} \rangle$ over this transition. The hyperfine levels of the ground state were well-resolved within the Doppler-broadened $D_{2}$ lines, but those of the excited states were not. We scanned the probe laser frequency over the Doppler-broadened resonance lines and derived $\langle F_{z} \rangle$ parameters from the integrated absorption coefficients. 
The largest $\langle F_{z} \rangle$ was 2.5 for the coated cell, corresponding to a degree of polarization of 0.63. We discuss spin flow during optical pumping, i.e., the flow from the incident photons to atoms mediated via photon absorption, from atoms to the vacuum via spontaneous emission, and from atoms to solid surfaces (cell walls) via surface collisions. The spin relaxation probabilities during single collisions with uncoated and coated surfaces are estimated by using this spin flow scenario. 

The remainder of the paper is organized as follows. The theoretical basis of the total spin measurement is described in Sec. \ref{sec:Theory}. In Sec. \ref{sec:Exp}, the experimental setup is presented. The results of polarization measurements on both uncoated and coated cells are described in Sec. \ref{sec:RandD}. The spin flow is discussed in the context of total spin measurement in Sec. \ref{sec:RandD}. The paper concludes with Sec. \ref{sec:Con}.

\section{Theory}\label{sec:Theory}
We present a method that monitors spin-dependent absorption. We use $^{133}$Cs atoms with a nuclear spin of 7/2 as an example, but the method is applicable to all alkali-metal atoms when the hyperfine levels of the ground state are optically resolved and the optical depth of the alkali-metal vapor is not excessive. We also assume in the following that the spatial distribution of atoms is uniform, which is the case in our experiments using buffer-gas-free alkali-metal vapor cells. The populations of the excited states are also neglected because light intensities are low, as discussed in Sec.~\ref{sec:RandD}.

In the low-intensity region where Beer's absorption law is applied, the light intensity $I_{\rm{out}}$ after transmission through an alkali-metal vapor cell is generally expressed as follows:
\begin{equation}
I_{\rm{out}} = I_{\rm{in}} \exp \left( -\frac{N}{V} \sigma L \right) \label{eq:transmition},
\end{equation}
where $I_{\rm{in}}$ is the light intensity prior to cell entry, $N$ is the number of atoms, $\sigma$ is the absorption cross-section, $L$ is the cell length, and $V$ is the cell volume. The measured $I_{\rm{out}}$ and $I_{\rm{in}}$ values yield $N\sigma L/V$, abbreviated $A$, as follows:
\begin{equation}
A = N \sigma \frac{L}{V} = - \ln \frac{I_{\rm{out}}}{I_{\rm{in}}} \label{eq:nsigmaL}.
\end{equation}

\begin{figure}[h]%
\centering
\includegraphics[width=0.9\textwidth]{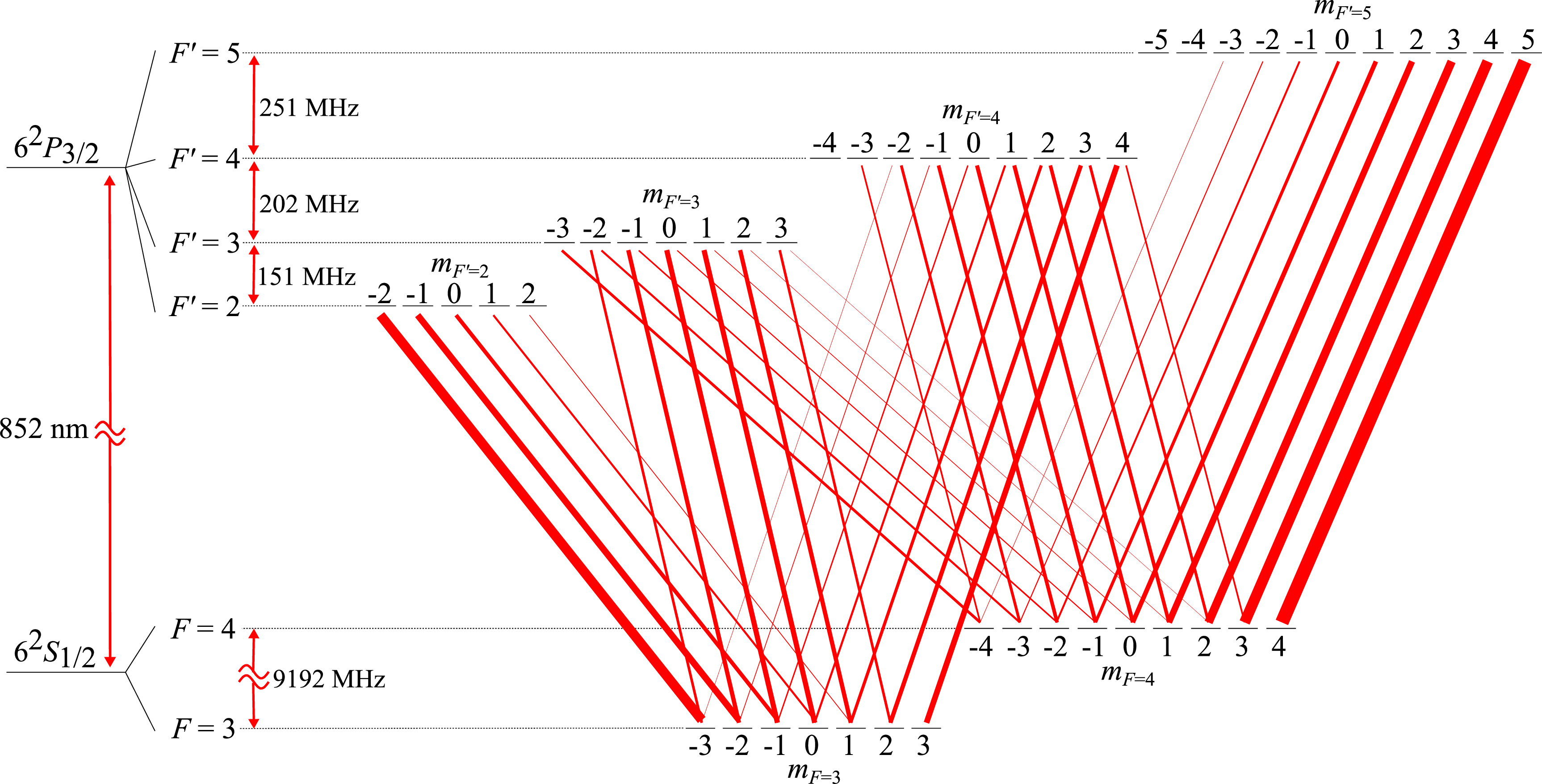}
\caption{The energy levels and transition strengths of a Cs $D_2$ line for $\sigma ^{+}$ polarization. The thickness values of the red lines indicate the transition strengths. The hyperfine interval of the ground state is an SI defining constant \cite{Ari77} and the excited intervals are referred to Ref. \cite{Ger03}.
}
\label{transition}
\end{figure}

Specifically, $\sigma$ in $A$ depends on the atomic state and the light frequency and polarization. Figure \ref{transition} shows the energy levels and transition strengths associated with the $D_{2}$ transition from the $^{2}S_{1/2}$ state to the $^{2}P_{3/2}$ state of Cs atoms. The thickness values of the red lines are proportional to the transition strengths when the incident light polarization is $\sigma^{+}$. 
We term the absorption cross-section $\sigma^{+(-)}_{F, m_{F} \to F'} $, which describes the transition from the lower level $|F, m_{F}\rangle$ to the upper level $|F^\prime, m_{F} +1 (m_{F}-1)\rangle$ by light with $ \sigma ^{+(-)} $ polarization. Here, $F$ is the quantum number of the total spin, and $m_F$ is the magnetic quantum number.
The transition strength for the electric dipole transition can be calculated as follows \cite{Met99}:
\begin{align}
\lvert \langle F^{\prime}, m_{F} \pm 1| d_{\pm 1} | F, m_{F} \rangle \rvert ^{2}
 = & (-1)^{ 2 \left( F^{ \prime} - m_{F} \pm 1  \right) } 
 \begin{pmatrix} F^{\prime} & 1 & F \\ - (m_{F} \pm 1) & \pm 1 & m_{F} \end{pmatrix}^{2} \nonumber\\
 &\times (-1)^{2 \left( J^{\prime}+I+F+1 \right)} (2F+1)(2F^{\prime}+1 )
\begin{Bmatrix} J^{\prime} & F^{\prime} & I \\ F & J & 1 \end{Bmatrix} ^{2} \nonumber \\
&\times  (-1)^{2 \left( L^{\prime}+S+J+1 \right)} (2J+1)(2J^{\prime}+1)
\begin{Bmatrix} L^{\prime} & J^{\prime} & S \\ J & L & 1 \end{Bmatrix} ^{2}
\lvert \langle L^{\prime} || d || L \rangle \rvert ^{2} \label{eq:transition_st}.
\end{align}
where $d_{\pm 1}$ is the spherical component of the dipole moment vector, $ \begin{pmatrix} F^{\prime} & 1 & F \\ - m_{F} \pm        1 & \pm 1 & m_{F} \end{pmatrix}$ is the 3$j$ symbol,  $\begin{Bmatrix} J^{\prime} & F^{\prime} & I \\ F & J & 1 \end{Bmatrix}$ and $\begin{Bmatrix} L^{\prime} & J^{\prime} & S \\ J & L & 1 \end{Bmatrix}$ are 6$j$ symbols, and $ \langle L^{\prime} || d || L \rangle $ is the reduced matrix element. For the Cs $D_2$ transition, $J=1/2$, $J^{\prime} = 3/2$, $L =0$, and $L^{\prime}=1$.

For each transition, the number of atoms in the lower level, i.e., $n_{F, m_{F} } \times N_{\rm{total}}$, must be considered when deriving $N$ in $A$. Here, $n_{F, m_{F} }$ and $N_{\rm{total}}$ are the atomic population in $|F, m_{F}\rangle$ and the total number of atoms in the cell, respectively. Note that $n_{F, m_{F} }$ is normalized to become $\sum_{F, m_{F}} n_{F, m_{F}} = 1.$

We also consider the frequency dependence. At room temperature, the Doppler-broadening of optical transitions [full width at half maximum (FWHM): 378 MHz] is much wider than both the natural linewidth of the Cs atoms (ca. 5 MHz) and the laser linewidth ($< 1~\rm{MHz}$ ). Thus, it is a good approximation, especially when the laser frequency is close to the transition frequency \cite{Hug18}, that the $A$ of each transition is weighted by $f( v = \frac{\Delta \omega_{F F'}}{\omega_{F F'}} c)$, derived from the velocity distribution function of the atoms $f(v)$. Here, $c$ is the speed of light, and $v=\frac{\Delta \omega_{F F'}}{\omega_{F F'}} c=\frac{\omega-\omega_{F F'}}{\omega_{F F'}} c$ is the velocity at which the light frequency that is Doppler-shifted from the original frequency $\omega$ matches the transition frequency $\omega_{F F'}$.
Note that we approximate $\Delta\omega_{FF'} \ll \omega_{FF'}$; $f(v)$ is normalized to $\int_{-\infty}^{\infty} f( v = \frac{\Delta \omega_{FF'}}{\omega_{FF'}} c ) d\omega =1$.

Thus, $A$ can be derived for the transition from the ground state $F = 3$ or $4$ to all excited states $F^{\prime} = 2,3,4$ or $3,4,5$ for light with a frequency bandwidth from $\omega$ to $\omega+d\omega$, and with $\sigma^{+(-)}$ polarization as follows: 
\begin{equation}
d A^{+(-)}_F (\omega)= \frac{L}{V} \sum_{m_{F}} \sum_{F'} n_{F, m_{F} } N_{\rm{total}} ~ f\left( v = \frac{\Delta \omega_{F F'}}{\omega_{F F'}} c \right)d \omega ~ \sigma_{F, m_{F}\to F'}^{+(-)} .
\end{equation}

By scanning the light frequency over the Doppler-broadened optical transitions to all excited $F^\prime$ levels, we obtain the integrated absorption coefficients for all transitions from $F$, as follows:

\begin{align}
 A^{+(-)}_F = \int d A^{+(-)}_F &= \frac{L}{V}  \sum_{F, m_{F}} n_{F, m_{F}} N_{\rm{total}}~\int f\left( v = \frac{\Delta \omega_{FF'}}{\omega_{FF'}} c \right) d \omega ~\sum_{F'}  \sigma_{F, m_{F}\to F'}^{+(-)}  \\
&= \frac{L}{V} \sum_{F, m_{F}} n_{F, m_{F}} N_{\rm{total}} ~ \sum_{F'}\sigma_{F, m_{F}\to F'}^{+(-)}. \label{eq:scanned_absprption}
\end{align}

\begin{table}[h]
\caption{The relative transition strengths of the Cs $D_{2}$ line using $\sigma^{\pm}$ polarized light. }
\label{transition_table}%
\renewcommand{\arraystretch}{1.7}
\begin{ruledtabular}
\begin{tabular}{rccr}
& Transition from $|F, m_{F} \rangle \to |F^\prime, m_{F^\prime}\rangle$ & Transition strength &\\
\hline
& $|3, m_{F} \rangle \to |2, m_{F} \pm 1 \rangle$    & $ 120 ~m_{F=3}^2 \mp 600 ~m_{F=3} + 720$ & \\
& $|3, m_{F} \rangle \to |3, m_{F} \pm 1 \rangle$     & $- \frac{315}{2} ~m_{F=3}^2 \mp \frac{315}{2} ~m_{F=3} +  1890$ & \\
& $|3, m_{F} \rangle \to |4, m_{F} \pm 1 \rangle$     & $ \frac{75}{2} ~m_{F=3}^2 \pm \frac{675}{2} ~m_{F=3} +  750$ &\\
& $|4, m_{F} \rangle \to |3, m_{F} \pm 1 \rangle$     & $ \frac{35}{2} ~m_{F=4}^2 \mp \frac{245}{2} ~m_{F=4} +  210$ &\\
& $|4, m_{F} \rangle \to |4, m_{F} \pm 1 \rangle$     & $ - \frac{147}{2} ~m_{F=4}^2 \mp \frac{147}{2} ~m_{F=4} +  1470$ &\\
& $|4, m_{F} \rangle \to |5, m_{F} \pm 1 \rangle$     & $ 56 ~m_{F=4}^2 \pm 616 ~m_{F=4} + 1680$ &\\
\end{tabular}
\end{ruledtabular}
\end{table}

Table \ref{transition_table} shows the strengths of the $D_{2}$ transition from the Cs atom $^{2}S_{1/2}$ state to the $^{2}P_{3/2}$ state, which is calculated by using Eq.~\eqref{eq:transition_st}. The $\pm$ and $\mp$ signs in the transition strengths reflect the $\sigma^{\pm}$ light polarization. 
From Table \ref{transition_table}, we obtain, for $F = 3$ and $F = 4$, 
\begin{align}
\sum_{F'=2, 3, 4}\sigma_{F=3, m_{F=3}\to F'}^{\pm} = a(\mp 420 ~m_{F=3} + 3360) , \label{eq:sigma_F3}\\
\sum_{F'=3, 4, 5}\sigma_{F=4, m_{F=4}\to F'}^{\pm} = a(\pm 420 ~m_{F=4} + 3360), \label{eq:sigma_F4}
\end{align}
where $a$ is a proportional constant.

This yields the integrated absorption coefficients for the transitions from the $F = 3$ and $F = 4$ levels: 
\begin{align}
A_{F = 3}^{\pm} =\frac{L}{V} a \sum_{F=3, m_{F=3}}n_{F=3, m_{F=3} } N_{\rm{total}} \left(  \mp 420 ~m_{F=3} + 3360 \right),\label{eq:absorption_F3}\\
A_{F = 4}^{\pm} = \frac{L}{V} a \sum_{F=4, m_{F=4}} n_{F=4, m_{F=4}} N_{\rm{total}} \left( \pm 420 ~m_{F=4} + 3360 \right). \label{eq:absorption_F4}
\end{align}

We next experimentally evaluate $\frac{L}{V}a N_{\rm{total}}$.
For Cs atoms in thermal equilibrium at room temperature, the populations $n_{F, m_{F}}$ are all described by $1/16$. Thus, the absorption coefficients of the Cs atoms at thermal equilibrium are as follows:
\begin{align}
\bar{A}_{F = 3}^{ \pm }  =\frac{L}{V} a \sum_{F=3, m_{F=3}} \frac{1}{16} N_{\rm{total}} \left( \mp 420 ~m_{F=3} + 3360 \right) = 1470~\frac{L}{V} a N_{\rm{total}}, \label{eq:absorption_F3_nopump}\\
\bar{A}_{F = 4}^{ \pm }  = \frac{L}{V} a \sum_{F=4, m_{F=4}} \frac{1}{16} N_{\rm{total}} \left( \pm 420 ~m_{F=4} + 3360 \right) = 1890~\frac{L}{V} a N_{\rm{total}}. \label{eq:absorption_F4_nopump}
\end{align}

From Eqs. \eqref{eq:absorption_F3}, \eqref{eq:absorption_F4}, \eqref{eq:absorption_F3_nopump}, and \eqref{eq:absorption_F4_nopump}, we obtain the expectation values for the spin $z$ components of the $F=3$ and $F=4$ states, thus
\begin{align}
\langle F_{z} \rangle  _{F}= \sum_{m_{F}} n_{F, m_{F}}  \times m_{F}, \label{eq:Fz}
\end{align}
as follows:
\begin{align}
\langle F_{z} \rangle_{F=3} = \sum_{m_{F=3}} n_{F=3, m_{F=3}}  \times m_{F=3} &= -\frac{7}{2} \frac{A_{F = 3}^{+} - A_{F = 3}^{-} }{\bar{A}_{F = 3}^{ + } + \bar{A}_{F = 3}^{- } },\\
\langle F_{z} \rangle_{F=4} = \sum_{m_{F=4}} n_{F=4, m_{F=4}}  \times m_{F=4} &= \frac{9}{2} \frac{A_{F = 4}^{+} - A_{F = 4}^{-} }{\bar{A}_{F = 4}^{+ } + \bar{A}_{F = 4}^{ - } }.
\end{align}

Finally, we obtain the expectation value of the $z$ component of the total spin angular momentum $\langle F_{z} \rangle$ as
\begin{align}
\langle F_{z} \rangle= \sum_{F, m_{F}} n_{F, m_{F}}  \times m_{F}=\langle F_{z} \rangle_{F=3}+ \langle F_{z} \rangle_{F = 4}.
\end{align}

The polarization of the total spin is
\begin{align}
P = \frac{ \langle F_{z} \rangle }{ \langle F_{z} \rangle_{\rm{max}} } = \frac{ \langle F_{z} \rangle }{ 4 },
\end{align}
where $\langle F_{z} \rangle_{\rm{max}}$ is the maximum $z$ component of the total spin.

Note that $\langle S_z \rangle$ can be derived using the following equation \cite{Bal75}:
\begin{align}
\langle S_z \rangle=\frac{1}{2I+1}(\langle F_{z} \rangle_{F=4}- \langle F_{z} \rangle_{F = 3}).
\end{align}

We can also obtain the populations of the hyperfine levels $n_{F}= \sum_{m_{F}} n_{F, m_{F}}$ ($F=3$ or $4$) from Eqs. \eqref{eq:absorption_F3}, \eqref{eq:absorption_F4}, \eqref{eq:absorption_F3_nopump} and \eqref{eq:absorption_F4_nopump} as follows:
\begin{align}
 n_{F=3}&= \sum_{m_{F=3}} n_{F=3, m_{F=3}} = \frac{7}{16} \frac{A_{F = 3}^{+} + A_{F = 3}^{-} }{\bar{A}_{F = 3}^{+ } + \bar{A}_{F = 3}^{- } },\\
 n_{F=4}&= \sum_{m_{F=4}} n_{F=4, m_{F=4}} = \frac{9}{16} \frac{A_{F = 4}^{+} + A_{F = 4}^{-} }{\bar{A}_{F = 4}^{ + } + \bar{A}_{F = 4}^{- } },
\end{align}
where 7/16 and 9/16, coefficients of relative populations of hyperfine levels in thermal equilibrium, are reasonably derived.

\section{Experimental setup} \label{sec:Exp}
\begin{figure}[tb]%
\centering
\includegraphics[width=0.9\textwidth]{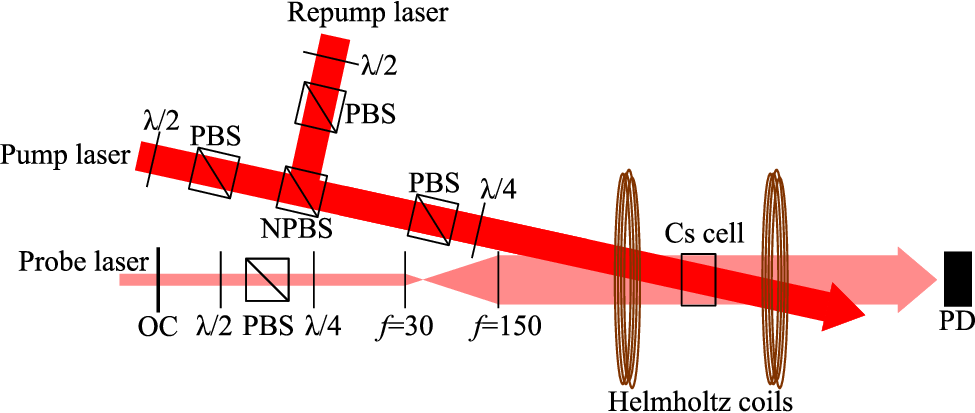}
\caption{Schematic diagram of the experimental setup. OC: optical chopper; $\lambda /4$: quarter waveplate; $\lambda /2$: half waveplate; PBS: polarizing beamsplitter; NPBS: non-polarizing beamsplitter; PD: photodiode; $f = 30$ and $f=150$: convex lenses with focal lengths of 30~mm and 150~mm, respectively (these expand the laser diameter).
}\label{setup}
\end{figure}

Figure \ref{setup} shows a schematic diagram of the experimental setup. We induced spin polarization in a Cs vapor cell employing optical pumping using a single pump laser beam or overlapping pump and repump laser beams at room temperature. 
Both the pump and repump beams exhibit Gaussian beam profiles. The pump beam diameter is 8 mm (1/$e^2$ diameter), and that of the repump beam is 9 mm. In most cases, the frequencies of the pump and repump lasers are tuned to resonances between $F = 4 \to F^{\prime} = 5$ and $F = 3 \to F^{\prime} = 4$, respectively. The lasers are frequency stabilized via polarization spectroscopy \cite{Har06} employing in-house servo circuits.

Spin polarization in the Cs vapor cell is monitored by the probe laser. The beam runs parallel to the axis of the cylindrical cell; the angle between the probe beam and the pump or repump beam is 5 degrees. The diameter of the probe laser beam, which also exhibites a Gaussian profile, is 12 mm (1/$e^2$ diameter). The probe laser frequency runs over the Doppler-broadened resonance lines from $F = 3$ to $F^{\prime}$ and from $F = 4$ to $F^{\prime}$. 
We modulate the probe laser intensity by using an optical chopper to perform lock-in measurements. The lock-in frequency is typically 2,540 Hz, i.e., much faster than the scan frequency of the probe laser (0.25 Hz). All experiments are performed eight times, and the data are averaged.

A magnetic field ($3~\rm{G}= 0.3~\rm{mT}$) is applied to the Cs cell (in parallel with the probe beam) by using Helmholtz coils. The magnetic field is low enough as compared to $A_{\rm{HF}} / \mu_B$. The Zeeman shifts of the magnetic sublevels are $<$ 5 MHz, and thus are much smaller than the Doppler broadening. The directions of the probe beam and the magnetic field defined the quantization axis ($z$ axis).

We use two types of Cs vapor cells: uncoated and anti-relaxation-coated cells. Both cells are cylindrical in shape and do not contain any buffer gas. The inner diameter and length of the uncoated cell are 27 and 17 mm, respectively, and they are 17 and 27 mm, respectively for the anti-relaxation-coated cell.
The anti-relaxation paraffin-coated cell is prepared in-house as described in Ref. \cite{Sek16}. The spin relaxation times are measured by using ``the relaxation in the dark'' method \cite{Fra59}, which employed optical rotation and resonance light \cite{Gra05}. Relaxation has two components as in Ref. \cite{Gra05}: a slow relaxation time of 73~ms and a fast relaxation time of 10~ms. According to Ref.~\cite{Gra05}, the former was attributed to uniform relaxation that reduces the polarization in both hyperfine levels, while the latter was attributed to electron randomization and spin-exchange collisions that transfer polarization from one level to another.

\section{Results and Discussion}\label{sec:RandD}

\subsection{Uncoated cell}\label{subsec:Uncoate}
\begin{figure}[tb]%
\centering
\includegraphics[width=0.95\textwidth]{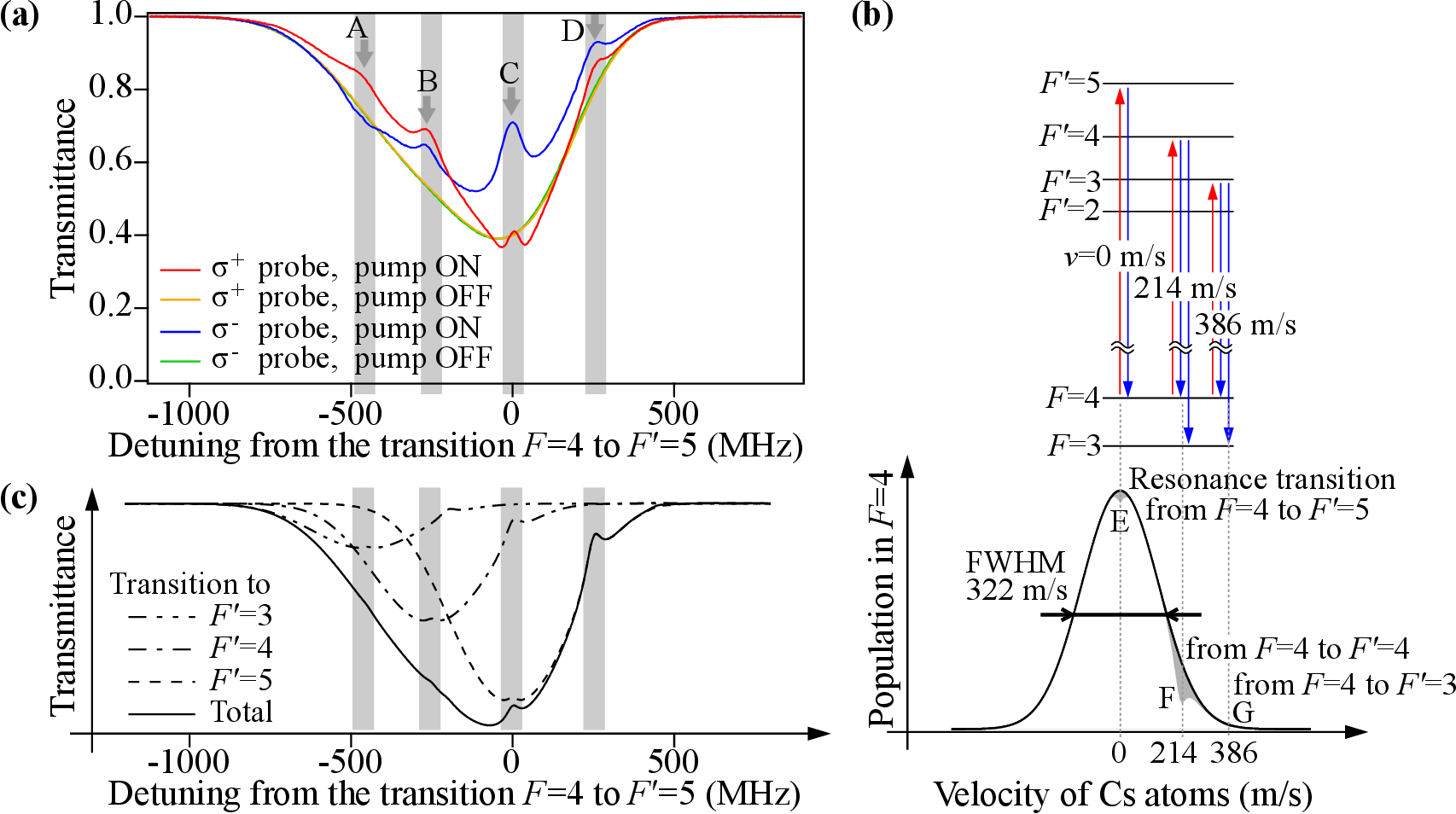}
\caption{(a) Transmittance spectra for the uncoated cell with the $\sigma^+$ pump laser ($F=4 \rightarrow F^{\prime}=5$) on and off. The spectra are derived using the $\sigma^+$ or $\sigma^-$ probe laser. The pump laser power is 4.4~mW (9.1~mW/cm$^2$), and the probe laser power is 9.4~$\mu$W (8.4~$\mu$W/cm$^2$). (b) The velocity distributions of Cs atoms in the $F$ = 4 state. The shaded areas schematically show population depletions. (c) Schematic decomposition of the transmittance spectrum into three Doppler-broadened resonance lines originated from the transitions from $F=4$ to $F^{\prime}=$3, 4, and 5.}
\label{231025Graph8}
\end{figure}

$\langle F_z \rangle$ is evaluated under each pumping condition by deriving eight integrated absorption coefficients calculated by using the respective transmittance spectra. 
Figure \ref{231025Graph8}(a) shows a typical set of four $I_{\rm{out}}/I_{\rm{in}}$ transmittance spectra for the uncoated cell. These are obtained by running the probe laser frequency over the transitions from the $F = 4$ hyperfine level. The other four absorption spectra are those for the transitions from the $F = 3$ hyperfine level. All four curves are recorded with the probe laser polarizations set to $\sigma^{+}$ and $\sigma^{-}$; the pump laser is either on or off. The pump and probe laser powers are 4.4~mW (intensity: 9.1~mW/cm$^2$) and $9.4~\rm{\mu W}$ (intensity: 8.4~$\mu$W/cm$^2$), respectively. The repump laser is not employed. The probe laser intensity is sufficiently low, as explained in more detail in the description of Fig.~\ref{231025Graph4} below.
The magnetic field is turned on and off in synchrony with the pump laser. The pump laser frequency is tuned to that of the transition from $F = 4 \to F^{\prime}=5$; the polarization is $\sigma ^{+}$.
With the pump laser off, Doppler-broadened resonance lines are apparent but the hyperfine levels of the upper state are not resolved. The spectral features are independent of probe polarization status. With the pump laser on, we observe structures created via optical pumping that depend on the velocities of Cs atoms along the laser beams \cite{Cha07}. 

\begin{figure}[tb]%
\centering
\includegraphics[width=0.7\textwidth]{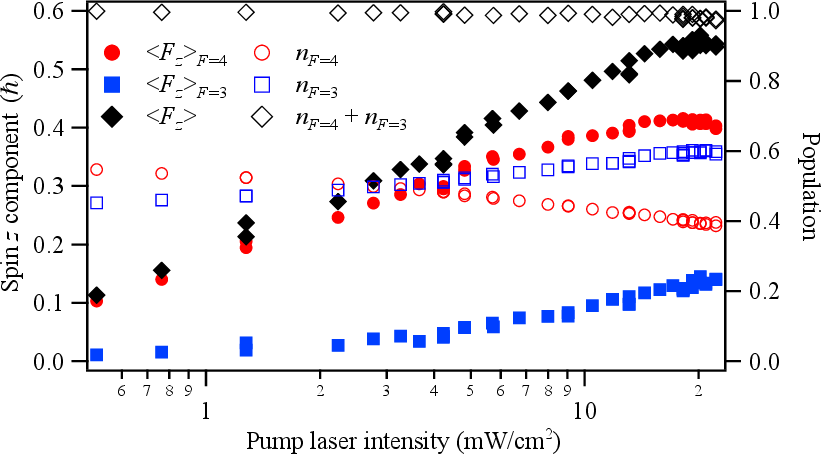}
\caption{Dependencies of the spin $z$ components and the level populations on the pump laser intensity delivered to the uncoated cell. The frequency of the $\sigma^+$ pump laser is tuned to that of the transition from $F = 4 \to F^{ \prime }= 5$. The probe laser power is 9.5~$\mu$W (8.5~$\mu$W/cm$^2$).}\label{231027Graph1}
\end{figure}

To aid qualitative understanding of the overall structure, Fig.~\ref{231025Graph8}(b) shows the velocity distribution of the Cs atoms in $F = 4$ when the pump laser is tuned to the transition from $F = 4$ to $F^\prime=5$. The shaded areas (E, F, and G) show the population depletions attributable to hyperfine pumping to the $F = 3$ level via the $F^{\prime} = 4$ and $F^{\prime} = 3$ levels and pumping to the excited states. We note that the atoms pumped to the $F=3$ level are measured with the transitions from $F=3$ level for evaluating $\langle F_{z} \rangle$, as described previously. The population in the excited states is not taken into account for evaluating $\langle F_{z} \rangle$ because its fraction is very small even at this relatively high pump intensity, as discussed below in the description of Fig. \ref{231027Graph1}. Peaks A to D in Fig.~\ref{231025Graph8}(a) reflect depletions of the $F=4$ atoms, as shown schematically in Fig.~\ref{231025Graph8}(c), as well as changes in the populations of the magnetic sublevels due to pump laser application. The changes in the sublevel populations also produce the difference between the spectra of the red ($\sigma^{+}$ probe polarization) and blue ($\sigma^{-}$ probe polarization) curves. For the transitions $F = 4$ to $F^\prime = 3$ and $F = 4$ to $F^\prime = 4$, the transition strengths of the $\sigma^{-}$ probe light are larger than those of the $\sigma^{+}$ probe light when $\langle F_z \rangle_{F=4} > 0$ (Table \ref{transition_table}). Such dependence on probe polarization is particularly evident at peaks A and B in Fig.~\ref{231025Graph8}(a). The transition from $F = 4$ to $F^{\prime}=5$ exhibits the opposite dependence on $\langle F_z \rangle_{F=4}$; such polarization dependence is particularly evident at peak C in Fig.~\ref{231025Graph8}(a).

It is interesting to note, as mentioned in Ref. \cite{Hug18}, that pumping using the $F=4$ to $F^{\prime}=3$ transition selects a much wider range of velocities than determined by the natural linewidth when the pump laser frequency is far from the transition frequency by the order of the Doppler broadening. Consideration of this subtle effect may be necessary to explain the structures of the transmittance spectra in more detail.

\begin{figure}[tb]%
\centering
\includegraphics[width=0.7\textwidth]{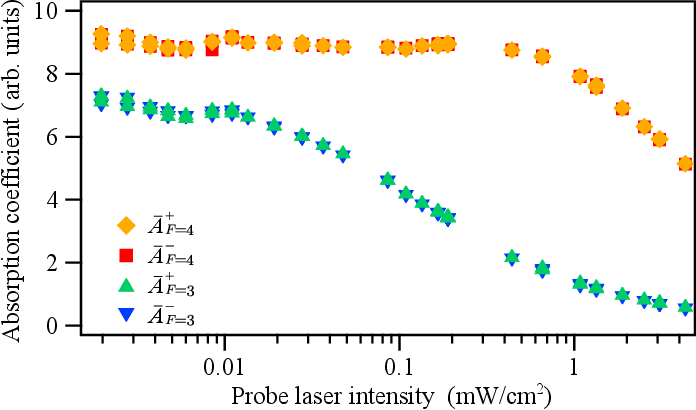}
\caption{Dependence of the integrated absorption coefficients on the probe laser intensity when the pump and repump lasers and the magnetic field are off for the uncoated cell.}
\label{231025Graph4}
\end{figure}

To investigate the concrete effects on the atomic polarization and population of the probe laser itself, we obtain the integrated absorption coefficients when the pump and repump lasers are off as functions of the probe laser power, as shown in Fig. \ref{231025Graph4}. We vary the probe laser power over a wide range; the horizontal axis of the figure is logarithmic. The four integrated absorption coefficients described by Eqs.~\eqref{eq:absorption_F3_nopump} and \eqref{eq:absorption_F4_nopump} are plotted. As the probe power increases, the integrated absorption coefficients fall because hyperfine pumping (by the probe laser) from the probed hyperfine level to the next level becomes faster than hyperfine relaxation, which is mainly caused by wall collisions. Note that this hyperfine pumping occurs at much lower light intensities than the saturation intensity of 1.1 mW/cm$^2$ for the $|F=4, m_F=4\rangle$ to $|F=5, m_F=5\rangle$ closed transition of the Cs atoms \cite{Met99}. In the power region below $10~\rm{\mu W}$ (8.9~$\mu$W/cm$^2$), the integrated absorption coefficients are independent of the power. The ratio of $\bar{A}^{\pm}_{F=4}$ to $\bar{A}^{\pm}_{F=3}$ is 9:7, in good agreement with the statistical weights of the hyperfine levels $F = 4$ and $F = 3$. Thus, in this power region, the power is sufficiently low to prevent optical pumping by the probe laser and to apply Beer's law; we thus hold the probe power below $10~\rm{\mu W}$ (8.9~$\mu$W/cm$^2$) in the following experiments. 

Next, we measure polarization produced by the pump laser only, i.e., without the repump laser, at a probe power of $9.5~\rm{\mu W}$ (8.5~$\mu$W/cm$^2$).
Figure \ref{231027Graph1} shows the spin $z$ components and the level populations as a function of the pump laser power. With increasing pump power, both $\langle F_{z} \rangle_{F=4}$ and $\langle F_{z} \rangle_{F=3}$ increase. However, in the high-power region, $\langle F_{z} \rangle_{F=4}$ rises less rapidly, and even begins to decrease because the population of the $F = 4$ level $n_{F=4}$ falls given the strong hyperfine pumping to the $F = 3$ level, as shown in Fig.~\ref{231027Graph1}. Thus, $\langle F_{z} \rangle $ becomes saturated in the high-power region. The highest degree of polarization $P$ is 0.55/4= 0.14. 
It is important to note that a slight decrease in the total population, $n_{F=4} + n_{F=3}$, in the high-power region is an indication of the population of the excited states. In other words, our measurements are performed in the low light intensity region where the population of the excited states is negligible.

\begin{figure}[tb]%
\centering
\includegraphics[width=0.7\textwidth]{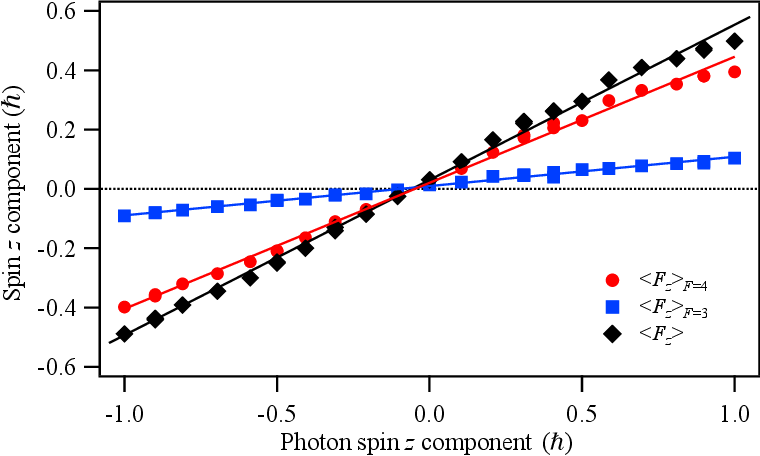}
\caption{Dependencies of the uncoated cell spin $z$ components on the spin $z$ component of the pump laser. The straight lines are linear data fits. The frequency of the pump laser is tuned to that of the transition from $F = 4 \to F^{ \prime }= 5$; the power is $4.5~\rm{mW}$ (9.3~mW/cm$^2$). The probe laser power is $9.6~\rm{\mu W}$ (8.6~$\mu$W/cm$^2$).}\label{231102Graph2}
\end{figure}

Third, the dependence of the spin $z$ components on the pump laser polarization (with the repump laser off) is measured, and is displayed in Fig. \ref{231102Graph2}. The horizontal axis is the photon spin component along the $z$ axis. The $\sigma^{+ \left( \rm{-} \right)}$ polarization corresponds to the spin component $1 (-1)$, but the elliptical polarizations change the spin components to between 1 and $-1$. The pump laser power is $4.5~\rm{mW}$ (9.3~mW/cm$^2$), and the probe power is $9.6~\rm{\mu W}$ (8.6~$\mu$W/cm$^2$). The $\langle F_{z} \rangle_{F=4}$, $\langle F_{z} \rangle_{F=3}$, and $\langle F_{z} \rangle$ are more-or-less proportional to the photon spin component. 

\begin{figure}[tb]%
\centering
\includegraphics[width=0.7\textwidth]{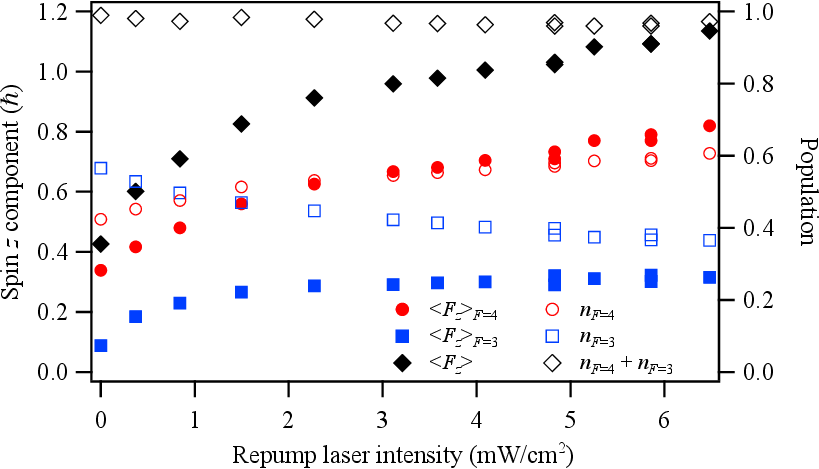}
\caption{Dependencies of the uncoated cell spin $z$ components and level populations on the repump laser intensity. The frequencies of the $\sigma^+$ pump and $\sigma^+$ repump lasers are tuned to those of the transitions from $F = 4 \to F^{ \prime }= 5$ and $F = 3 \to F^{ \prime }= 4$, respectively. The pump laser power is $4.2~\rm{mW}$ (8.7~mW/cm$^2$), and the probe laser power is $1.9~\rm{ \mu W}$ (1.7~$\mu$W/cm$^2$).}
\label{230202Graph2_4}
\end{figure}

To explore the effects of hyperfine laser repumping, we finally measure $\langle F_z \rangle$ by using both the pump and repump lasers. The repump laser is tuned to the transition from the $F = 3$ level to the $F^{\prime} = 4$ level. Figure~\ref{230202Graph2_4} shows the repump power dependency of $\langle F_z \rangle$. The pump laser power is $4.2~\rm{mW}$ (8.7~mW/cm$^2$), and the probe laser power is $1.9~\rm{ \mu W}$ (1.7~$\mu$W/cm$^2$). $\langle F_{z} \rangle $ monotonically increased with the repump power, attributable principally to a repumping-induced increase in the population of the $F = 4$ level. The highest degree of polarization $P$ apparent in Fig.~\ref{230202Graph2_4} is $1.2/4= 0.30$, which is three-fold that at 0~mW repump power. 

\subsection{Coated cell}\label{subsec:Coate}
Polarization of alkali-metal atoms in uncoated vapor cells is randomized by atomic collisions with the cell walls \cite{Bal75}. Spin relaxation times are thus generally determined by deriving the mean travel times between the walls; this is $7\times10^{-5}$~s in the uncoated cell discussed in the previous section.
In contrast, anti-relaxation-coated cells exhibit much longer spin relaxation times because appropriate coatings can prevent spin relaxation if $10^4$ collisions (or even more) occur \cite{WuZ21}. The coated cell used in this experiment exhibited relaxation times of the order of 10~ms, as mentioned in Sec. \ref{sec:Exp}.
The velocities of alkali-metal atoms in coated cells change after collisions with the walls and residual background gas molecules but the atoms retain their spin polarization \cite{Sek16}; all atoms across the entire velocity distribution become spin-polarized. Thus, spin polarization is more pronounced in coated than uncoated cells. However, possible strong hyperfine pumping must be considered. 

\begin{figure}[tb]%
\centering
\includegraphics[width=0.7\textwidth]{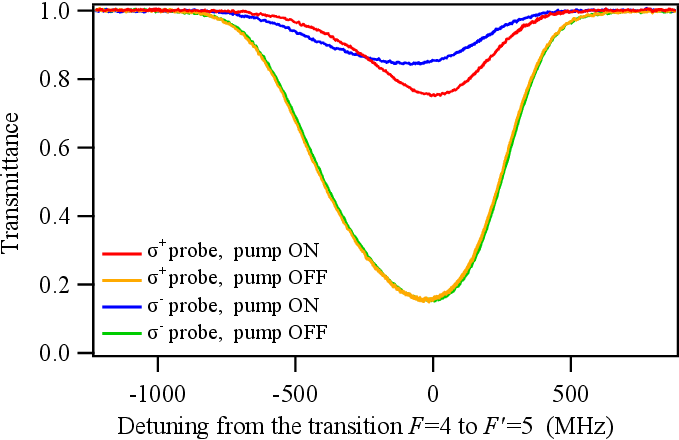}
\caption{Transmittance spectra for the coated cell with the $\sigma^+$ pump laser ($F=4 \rightarrow F^{\prime}=5$) on and off, measured using the $\sigma^+$ or $\sigma^-$ probe laser. The pump laser power is 23~$\mu$W (48~$\mu$W/cm$^2$) and the probe laser power is 40~nW ($36~\rm{ n W / cm^2}$).}
\label{240322Graph10}
\end{figure}

The transmittance spectra in Fig. \ref{240322Graph10} reveals just such different optical pumping properties of coated cells compared to those of uncoated cells. The powers of the pump and probe lasers are 23~$\mu$W (48~$\mu$W/cm$^2$) and 40~nW ($36~\rm{ n W / cm^2}$), respectively. The transmittances with the pump laser on are much higher than in the uncoated cell because hyperfine pumping is stronger in the coated cell. In addition, and in contrast to what Fig.~\ref{231025Graph8} shows for the uncoated cell, the Doppler-broadened absorption lines of the coated cell lack peaks because the optically pumped atoms are distributed across the entire velocity range when the velocities are changed by collisions with the wall and background gas. 

As we did with the uncoated cell, we first determined the probe power range over which the probe laser does not affect the integrated absorption coefficients. The top of this power range is under 50~nW ($45~\rm{ n W / cm^2}$), thus is much lower than the 10~$\mu$W ($8.9~\rm{ \mu W / cm^2}$) for the uncoated cell. 

\begin{figure}[tb]%
\centering
\includegraphics[width=0.7\textwidth]{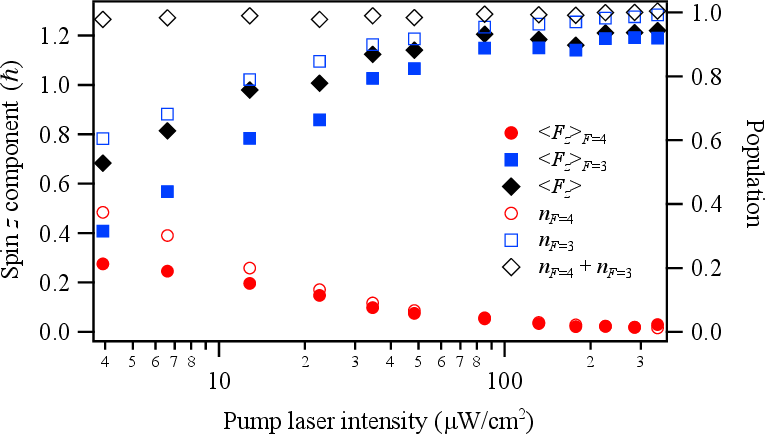}
\caption{Dependencies of the spin $z$ components and the populations on the $\sigma^{+}$ pump laser intensity for the coated cell. The frequency of the pump laser is tuned to that of the transition from $F = 4 \to F^{ \prime }= 5$. The probe laser power is $40~\rm{ n W}$ ($36~\rm{ n W / cm^2}$).}\label{240322Graph3}
\end{figure}

Figure~\ref{240322Graph3} shows the pump power dependence of the spin $z$ components. As in the uncoated cell, $\langle F_{z} \rangle$ increases with increasing pump power. However, it becomes saturated around $40~\rm{\mu W}$  ($83~\rm{ \mu W / cm^2}$), which is much lower than in the uncoated cell, reflecting the strong hyperfine pumping in the coated cell. The $F = 4$ population is almost completely depleted at a pump power over $40~\rm{\mu W}$ ($83~\rm{ \mu W / cm^2}$). The saturated value of $\langle F_{z} \rangle$ is $1.2$, 2.2-fold times higher than that of the uncoated cell. 

\begin{figure}[tb]%
\centering
\includegraphics[width=0.7\textwidth]{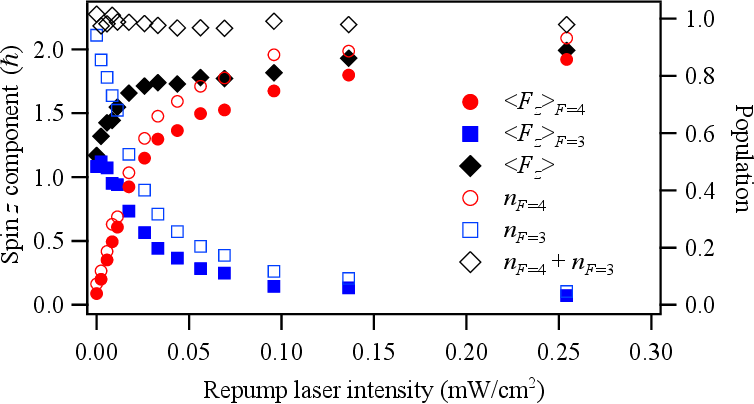}
\caption{Dependencies of the spin $z$ components and the populations on the repump laser intensity for the coated cell. The frequencies of the pump and repump laser are tuned to those of the transitions from $F = 4 \to F^{ \prime }= 5$ and $F = 3 \to F^{ \prime }= 4$, respectively. The pump laser power is $21~\rm{\mu W}$(43 $\mu$W/cm$^2$), and the probe laser power is $40~\rm{ n W}$ ($36~\rm{ n W / cm^2}$).}
\label{240404Graph1}
\end{figure}

Figure~\ref{240404Graph1} shows the dependencies on the repump laser power. The pump laser power is 21~$\mu$W (43~$\mu$W/cm$^2$). 
$\langle F_{z} \rangle$ becomes saturated around 0.09~mW (0.14~mW/cm$^2$); the saturation value is much higher than when repumping is off, as shown in Fig. \ref{240322Graph3}. The saturated $\langle F_z \rangle$ is 2.0, corresponding to $P = 0.50$. When repumping is off, $\langle F_{z} \rangle _{F=3}$ dominated $\langle F_{z} \rangle$, but $\langle F_{z} \rangle _{F=4}$ is dominant when repumping is on because of strong hyperfine pumping to the $F = 3$ level by the pump laser and hyperfine repumping to the $F = 4$ level by the repump laser.

To increase polarization, we investigate all pump and repump transition combinations. The highest polarization is $\langle F_{z} \rangle \sim 2.5$ and $P \sim 0.63$, achieved via $F = 4 \to F^{\prime} = 3$ pumping at 0.07~mW  ($0.14~\rm{ m W / cm^2}$) and $F = 3 \to F^{\prime} = 3$ repumping at 2.5~mW ($4.2~\rm{ m W / cm^2}$). A polarization of close to 1 may be obtained using other configurations, such as $D_1$ pumping, but we reserve such work for the future.

As discussed above, the dependences of $\langle F_{z} \rangle$ evaluated with the proposed method on pump or repump laser parameters, such as intensity, polarization, and frequency, are found to be in reasonable agreement with qualitative considerations for both uncoated and coated cells. However, the validity of the method should ultimately be examined using an alternative experimental approach. In this regard, we are planning to evaluate $\langle F_{z} \rangle$ mechanically using a Cs vapor cell hung in a torsion pendulum \cite{Yas21} on the basis of the Einstein-de Haas effect \cite{Ein15}. 
Quantitative reproduction of experimental results for theoretical verification requires modeling of optical pumping processes, including various types of relaxation caused by collisions with other atoms and cell walls, radiation trapping, and velocity changing collisions. The resulrs found in Refs.~\cite{Gra05, Sel08, Roc10} and a package for Mathematica \cite{RocHP} will be useful for the calculation.

\subsection{Spin flow}\label{balance}
\begin{figure}[tb]%
\centering
\includegraphics[width=0.7\textwidth]{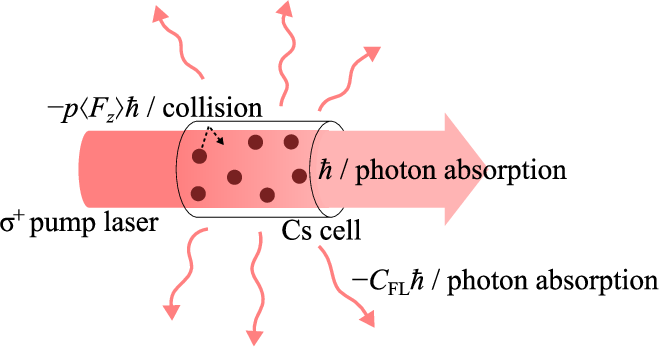}
\caption{Spin flow during optical pumping through the cell. The atoms receive an angular momentum of $\hbar$ per absorbed photon from the pump laser. The atoms lose an angular momentum of $C_{\rm{FL}} \hbar$ per single absorbed photon via spontaneous emission and an angular momentum of $p \langle F_{z} \rangle \hbar$ during each collision with the cell wall.}
\label{spin_flow}
\end{figure}

We determine the $\langle F_z \rangle$ values of the total angular momentum stored in uncoated and coated Cs cells. This enables us to introduce a simple spin flow model to describe the optical pumping process. In the steady states, the spin input from the pump laser to Cs atoms must be balanced with the spin output from the atoms to the environment. We consider two main features of the spin output, as schematically shown in Fig.~\ref{spin_flow}. The first is spontaneous emission of atoms, and the second is spin relaxation caused by collisions with the cell wall. Hence, the following equation is valid:

\begin{align}
R_{\rm{abs} } \hbar = C_{\rm{FL}}R_{\rm{abs} } \hbar + p \Gamma \langle F_{z} \rangle N_{\rm{total}} \hbar \label{eq:balance2_1},
\end{align}
where $R_{\rm{abs}}$ is the rate of photon absorption from the $\sigma^+$ pump laser, $C_{\rm{FL}}$ is the average spin emission factor per absorbing photon, $p$ is the spin relaxation probability for each wall collision, and $\Gamma$ is the wall collision rate of a single atom. $C_{\rm{FL}} = 0$ when unpolarized light is emitted, and $C_{\rm{FL}} = 1 (-1)$ when $\sigma^{+(-)}$ light is emitted.

$N_{\rm{total}}$ is calculated from the atomic density and the volume of the cell. The atomic density can be derived from the transmittance spectra. The integral of the absorption cross section $\sigma(\omega)$ associated with the transition $i$ to $k$ is
\begin{align}
\int_{0}^{\infty} \sigma{\left( \omega \right)}~d\omega = 2 \pi^2 r_{0} c f_{ik}, \label{eq:CS}
\end{align}
where $\omega$ is the angular frequency of light, $r_{0}$ is the classical electron radius, and $f_{ik}$ is the oscillator strength of the transition from $i$ to $k$. $r_{0}$ is $2.82 \times 10^{-15}~\rm{m}$ \cite{Cor78}, and $f_{ik}$ of the Cs $D_{2}$ line is 0.7149 \cite{Mor00}. 

The estimated atomic densities of our uncoated and coated cells are $3.4 \times 10^{10}~\rm{cm^{-3}}$, respectively. The measured atomic densities are consistent with the atomic density of $3.9 \times 10^{10}~\rm{cm^{-3}}$ derived from the Cs vapor pressure \cite{Alc84} at room temperature. 

The spin input parameters and those of the spin outputs under certain pumping conditions are shown in Table \ref{tab:balance}. $R_{\rm{abs}}$ is determined by using the absorbed pump laser power. The power inputs into the Cs atoms are $1.5~\rm{mW}$ and $4.4~\rm{\mu W}$ for the uncoated and coated cells, and 58\% and 73\% of the powers are absorbed by the Cs atoms, respectively. Note that the repump laser is off. $C_{\rm{FL}}$ depends on the spin states of the atoms and the optical transitions; these are roughly evaluated by assuming linear population distributions among the magnetic sublevels of the $F = 4$ level; this ensures that the observed $\langle F_z \rangle_{F = 4}$ is reproduced. We consider only the $F = 4$ to $F^{\prime} = 5$ transition during pumping. $\Gamma$ is estimated from the cell dimensions.

We derive $p$ from the other parameters using the balance relation of Eq. \eqref{eq:balance2_1}. $p$ is 1.6 for the uncoated cell. Considering an uncertainty of 10–20\% for each estimated parameter and processes that are not included in the simple spin flow model, such as radiation trapping of emitted light, we believe that this value is consistent with 1, and therefore $\bm{F}$ is randomized with a probability close to 1 by a single collision with uncoated glass surfaces \cite{Bal75, Sek18}. Turning to the anti-relaxation-coated cell, $p$ is $4 \times 10^{-3}$; $\bm{F}$ is completely randomized after about 300 wall collisions. The spin relaxation time $(p \Gamma)^{-1}$ is $2 \times 10$~ms, consistent with the relaxation time data (Sec. \ref{sec:Exp}).

\begin{table}[tb]
\caption{Estimated spin flow parameters. The frequency of the $\sigma^{+}$ pump laser is tuned to that of the transition from $F = 4 \to F^{ \prime }= 5$. The pump power inputs into the Cs atoms are $1.5 ~\rm{mW}$  ($3.1~\rm{ m W / cm^2}$) and $4.4~\rm{\mu W}$  ($9.1~\rm{ \mu W / cm^2}$) for the uncoated and coated cells, respectively. }
\label{tab:balance}
\begin{ruledtabular}
\begin{tabular}{ccccccc|c}
					& $\langle F_{z} \rangle _ {F=4}$ & $R_{\rm{abs} }\rm{(s^{-1})}$ & $C_{\rm{FL}}$ & $\Gamma \rm{(s^{-1})}$ & $\langle F_{z} \rangle$ & $ N_{\rm{total}}$ & $p $						\\ \hline
Uncoated cell  & 0.29 &$3.7 \times 10^{15}$ & $ 0.26 $ & $1.5\times 10^{4}$ & 0.34 & $3.3 \times 10^{11}$ &1.6\\
Coated cell  & 0.21 &$1.4 \times 10^{13}$ & $ 0.24 $ & $1.6 \times 10^{4}$ & 0.89 & $2.1 \times 10^{11}$ & $4 \times 10^{-3}$\\
\end{tabular}
\end{ruledtabular}
\end{table}

\section{Conclusions} \label{sec:Con}
We used absorption monitoring under a low magnetic field to measure $\langle F_z \rangle$, i.e., the expectation values of the quantization ($z$) axis components of the total angular momentum of Cs atoms in the electronic ground state in uncoated and anti-relaxation-coated vacuum cells at room temperature. Our method was based on the differences in the optical transition strengths among magnetic sublevels, and required only well-resolved hyperfine splitting data for the ground state. Transmittance spectra were obtained by running the probe laser frequency over the Doppler-broadened $D_2$ lines. We derived $\langle F_{z} \rangle$ values from the integrated absorption coefficients. The largest $\langle F_{z} \rangle$ was 2.5 for the coated cell, corresponding to a degree of polarization ($P$) of 0.63. The spin relaxation probabilities per wall collision, which were $\sim1$ for the uncoated cell and $4\times 10^{-3}$ for the coated cell, were estimated using a simple model of spin flow through cells subjected to optical pumping. We plan to further study spin transfer from atoms to other systems (such as cell surfaces).

\begin{acknowledgments}
This work was supported by JSPS KAKENHI Grant No. JP17H02933, a 2022 Sasakawa Scientific Research Grant No. 2022-2017 from the Japan Science Society, and a 2023 Research Grant (C) No. 2237011 from the Tateisi Science and Technology Foundation. WK and JWM acknowledge the support of the Canada Foundation for Innovation; the Canada Research Chairs program; and the Natural Sciences and Engineering Research Council of Canada (NSERC) SAPPJ-2016-00024, SAPPJ-2019-00031 and SAPPJ-2023-00029.
\end{acknowledgments}

\providecommand{\noopsort}[1]{}\providecommand{\singleletter}[1]{#1}%

\end{document}